\documentclass[preprints,briefreport,accept,pdftex,moreauthors,pdftex]{Definitions/mdpi} 

\usepackage{bm}

\usepackage{changes}
\definechangesauthor[color=orange]{Jan}
\definechangesauthor[color=green!50!black]{Wim}
\definechangesauthor[color=blue]{rev}
\firstpage{1} 
\pubvolume{1}
\issuenum{1}
\articlenumber{0}
\pubyear{2022}
\copyrightyear{2022}
\datereceived{} 
\dateaccepted{} 
\datepublished{} 
\hreflink{https://doi.org/} 

\Title{Nuclear $\text{C}(e,e'p)$ Transparencies in a Relativistic Glauber Model}

\TitleCitation{Nuclear $\text{C}(e,e'p)$ Transparencies in a Relativistic Glauber Model}

\Author{Wim Cosyn 
 $^{1,2,}$*\orcidA{} and Jan Ryckebusch $^{2,}$*\orcidB{}}

\AuthorNames{Firstname Lastname, Firstname Lastname and Firstname Lastname}

\AuthorCitation{Cosyn, W.; Ryckebusch, J.}

\address{%
$^{1}$ \quad Department of Physics, Florida International University,  Miami, FL 33199, USA\\
$^{2}$ \quad Department of Physics and Astronomy, Ghent University, B9000 Gent, Belgium}

\corres{Correspondence: wcosyn@fiu.edu (W.C.); jan.ryckebusch@ugent.be (J.R.)}

\abstract{In light of the recent Jefferson Laboratory (JLab)
data for the nuclear $^{12}\text{C}(e,e'p)$ transparencies, 
calculations, obtained in a relativistic multiple scattering Glauber approximation, are discussed.  
The
shell-separated $^{12}$C transparencies are shown 
and it is concluded
 that the $p$-shell nucleons are 75\% more transparent than the $s$-shell ones.  The presented comparisons between 
the calculations made here and the current 
$^{12}\text{C}(e,e'p)$ data show no clear indication for the onset of color transparency when implemented within the color diffusion model with standard parameters.}


\begin{document}

Forty years after it was first predicted in high-energy QCD~\cite{Mueller:1982bq, Brodsky:1982kg}, we still do not have a full understanding of the 
color transparency (CT) phenomenon in the intermediate (few to ten GeV) energy regime;
 for a recent review, see~\cite{Dutta:2012ii}. 
In reactions with high momentum transfer, CT predicts the formation of hadrons in small-sized configurations.  Due to the small transverse size, color multipoles vanish and final (or initial) state interactions of the hadron with a 
nuclear ($A$) 
medium get very strongly suppressed.  This phenomenon can be quantified in the nuclear transparency 
observable, 
\begin{equation}
 T=\frac{\mathrm{d}\sigma^A}{\mathrm{d}\sigma^A_\text{PWIA}}\;,   
\end{equation}
being the ratio of the cross sections for an exclusive nuclear knockout process in kinematic conditions approaching quasi-free production to the same 
process in the plane wave impulse approximation (PWIA) limit of no final-state interactions (FSI) of the knocked out particles with the nuclear 
medium.  The PWIA denominator has theoretical assumptions for the cross section entering; 
see, e.g.,~\cite{HallC:2020ijh} for more details.  
The color transparent limit in which the nuclear opacity gets fully suppressed by QCD mechanisms is reached for $T \to 1$.  At low and medium resolution 
scales, the opaqueness of the nuclear medium for hadron transmission results in $T<1$.  A fundamental question is at what resolution scales the CT phenomenon starts to suppress the nuclear opaqueness to a sizable degree.  Mapping this transition can teach us more about the transition from hadronic to partonic degrees of freedom and is of importance to the extraction of non-perturbative partonic distribution functions as CT plays a role in the QCD factorization theorems.  

Nuclear transparencies for $(e,e'p)$ reactions have been measured by a range of experiments~\cite{Garino:1992ca,O'Neill:1994mg,Makins:1994mm,Garrow:2001di,Rohe:2005vc,Abbott:1997bc,Dutta:2003yt,CLAS:2012usg,CLAS:2018xsb,HallC:2020ijh}.  Most recently, 
the Jefferson Laboratory  
(JLab) 12~GeV Hall C measurement~\cite{HallC:2020ijh} extended the kinematic range up to 
four-momentum transfers 
squared, 
$Q^2=14~\text{GeV}^2$, and observed no direct indications for the onset of CT when comparing the data to theoretical predictions.  On the other hand, 
proton transparencies measured earlier in $^{12}\text{C}(p,2p)$~\cite{Leksanov:2001ui} show an increase for
$Q^2$
between 5 and 10~GeV$^2$, followed by a decrease of $T$ up to 14 GeV$^2$.   For meson knockout reactions, signs of the onset of CT have been observed 
for the pion~\cite{Clasie:2007gq} and $\rho^0$~mesons~\cite{ElFassi:2012nr} at JLab. 
In this brief report,
results of
the 
transparency calculations for the $^{12}C(e,e'p)$ reaction are given. 
Earlier results by us of 
 the  relativistic multiple scattering Glauber approximation (RMSGA) 
 calculations for the total nuclear transparencies 
are summarised,
and
new calculations for the shell-separated transparencies are included. The Hall C Collaboration 
will report data for the $^{12}C(e,e'p)$ 
transparencies for separate shells in the near future.  First, let us start with a short overview of the formalism.

The
framework of 
RMSGA~\cite{Ryckebusch:2003fc} is used.  In the RMSGA, small-angle scattering of particles with momenta of a few 100 MeV and higher (so that 
its wavelength 
is small compared to the interaction range) is implemented in the eikonal approximation.  In a nuclear knockout reaction, the plane wave of the hit proton with shell-model quantum numbers,
 $\alpha_i$, in the initial state, acquires a Glauber eikonal phase in the photon-nucleon collision point, 
$(\bm b,z)$:

\begin{equation}
     \mathcal{G}(\bm b,z)=
 \prod _{\alpha \ne \alpha_{i}} \int d^3 r'
\left| \phi _ {\alpha} \left( \bm r'     \right) \right|^2
\left[
1 -  
  \theta \left( z - z' \right) \Gamma \left(
\bm b' -
\bm b \right) \right]\, .\label{eq:G}
\end{equation}
Here,
the $z$-direction is along the momentum of the ejected particle and $\bm b$ is the associated impact parameter.  
The Heaviside function, $\theta(z-z')$, ensures that only the nuclear medium in the forward direction contributes to the nuclear opaqueness, and 
the 
integral over the nuclear volume, $\int d^3 r'$ is weighted with the single-particle densities of the remaining nucleons.  The product,
 $\prod 
_{\alpha \ne \alpha_{i}}$, takes care of the full multiple scattering series.  The profile function, $\Gamma$, adopts a Gaussian 
form:
\begin{equation}
\label{eq:gamma}
\Gamma (\bm b) =
\frac{\sigma^{\text{tot}}(1-i\epsilon)}
{4\pi\beta^2}\exp{\left(-\frac{\bm b^2}{2\beta^2}\right)}\,, 
\end{equation}
where the total cross section, $\sigma^{\text{tot}}$, slope parameter $\beta$, and the ratio $\epsilon$ of the imaginary to real part of the nucleon-nucleon ($NN$) scattering amplitude, 
depend on the momentum of the ejected 
hadron and are 
parameterized using $NN$  
scattering data.  The eikonal phase encodes the effects of the FSI between the knocked out particle and the nuclear remnant (absorption, rescattering). The color transparency effect can be included in the profile function of Equation~(\ref{eq:gamma}) by replacing the total cross section parameter with a position dependent one using the quantum diffusion model of Refs.~\cite{Farrar:1988me,Frankfurt:1994kt}.  In the quantum diffusion model, the cross section evolves from a reduced value (reflecting the small-sized color transparent configuration) to its normal value along a coherence 
length, $l_c$:
\begin{equation}
\sigma^{\text{tot}} \longrightarrow
{ \sigma^{\text{eff}}(z) } =   
{ \sigma^{\text{tot}}} \biggl\{ \biggl[
 \frac{z}{l_c} +
 \frac{<n^2 k_t^2>}{Q^2} \left( 1- \frac{z}{l_c} 
\right) \biggr]
\theta(l_c-z) + 
\theta(z-l_c) \biggr\} \,  \; .
\label{eq:diffusion}
\end{equation}
Here, $l_c=2p/\Delta M^2$, with $p$ the momentum of the knocked-out 
proton and $\Delta M^2$ the difference in mass squared between the point-like configuration and the normal-sized nucleon.
 For the parameters in $\sigma^{\text{eff}}$, 
$\Delta M^2=1.0~\text{GeV}^2$,  $n=3$ reflects the elementary constituents in the proton, and $k_t=0.35~\text{GeV}$ 
the average transverse momentum of a quark inside a hadron~\cite{Farrar:1988me}.

For the $A(e,e'p)$ reaction, the RMSGA is implemented in an unfactorized manner, with the eikonal phase entering in the amplitude.  Kinematics and dynamics are treated relativistically, using mean-field nuclear wave function computed in the Serot-Walecka model~\cite{Furnstahl:1996wv}.  The calculation of the differential cross section entering in the transparency includes a summation over the mean-field quantum numbers of the final state residual $A-1$ target remnant.  In RMSGA, this summation corresponds to a summation over the single-particle quantum numbers of the bound nucleon in the initial nucleus before it interacts with the virtual photon.  The shell-separated transparencies shown below are obtained by limiting this summation to the quantum numbers of only the 
$s$-shell 
or $p$-shell protons in $^{12}$C; 
see Refs.~\cite{Ryckebusch:2003fc,Lava:2004zi} for more details on the formalism.  The effect of short-range correlations in the FSI can be 
accommodated 
by modifying the density, appearing in Equation~(\ref{eq:G}), but this does not generate appreciable differences in the final result; 
for details, 
see
Ref.~\cite{Cosyn:2013qe}. 

\begin{figure}[H]
\includegraphics[width=8 cm]{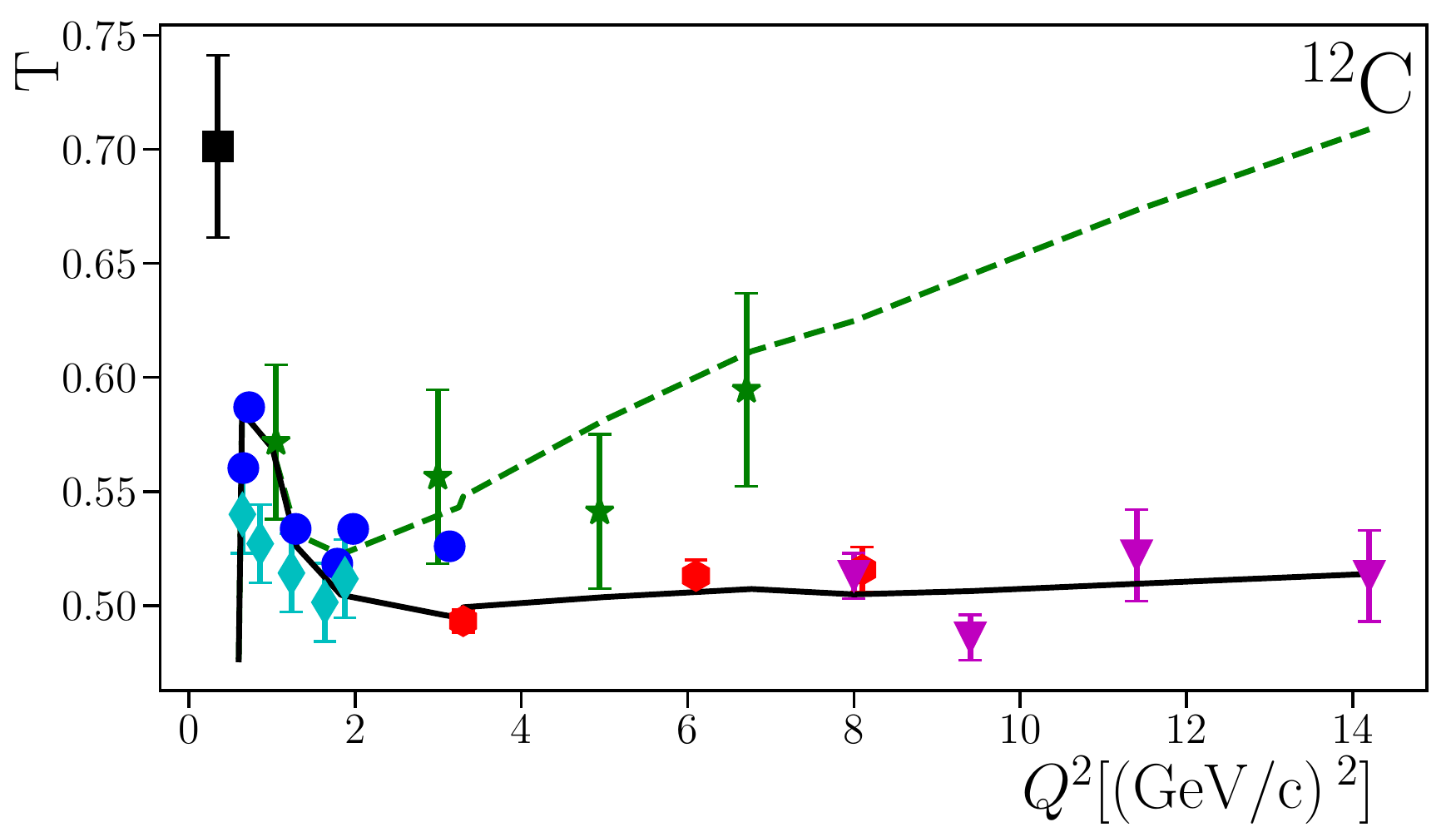}
\caption{Computed $^{12}$C$(e,e'p)$ nuclear transparencies versus the 4-momentum transfer squared, 
$Q^2$, in quasielastic kinematics.  The solid black line represents 
the 
relativistic multiple scatternig Glauber approximation 
(RMSGA) 
calculation that excludes color transparency (CT)
effects.  The dashed green line
  shows the RMSGA calculation including the CT effects with the prescription of Equation~(\ref{eq:diffusion}).  Data
  are from Refs.~\cite{Garino:1992ca} (black squares),
  \cite{O'Neill:1994mg,Makins:1994mm} (red hexagons),
  \cite{Garrow:2001di} (green stars), \cite{Rohe:2005vc} (cyan
  diamonds), \cite{Abbott:1997bc,Dutta:2003yt} (blue circles) and~\cite{HallC:2020ijh} (magenta triangles).  Data
  and calculations do not include the $c_A=1.11$ correlation factor, often applied in the transparency ratio~\cite{Lava:2004zi}. 
\label{fig:carbon_data}
}
\end{figure}

\begin{figure}[H]
\includegraphics[width=8 cm]{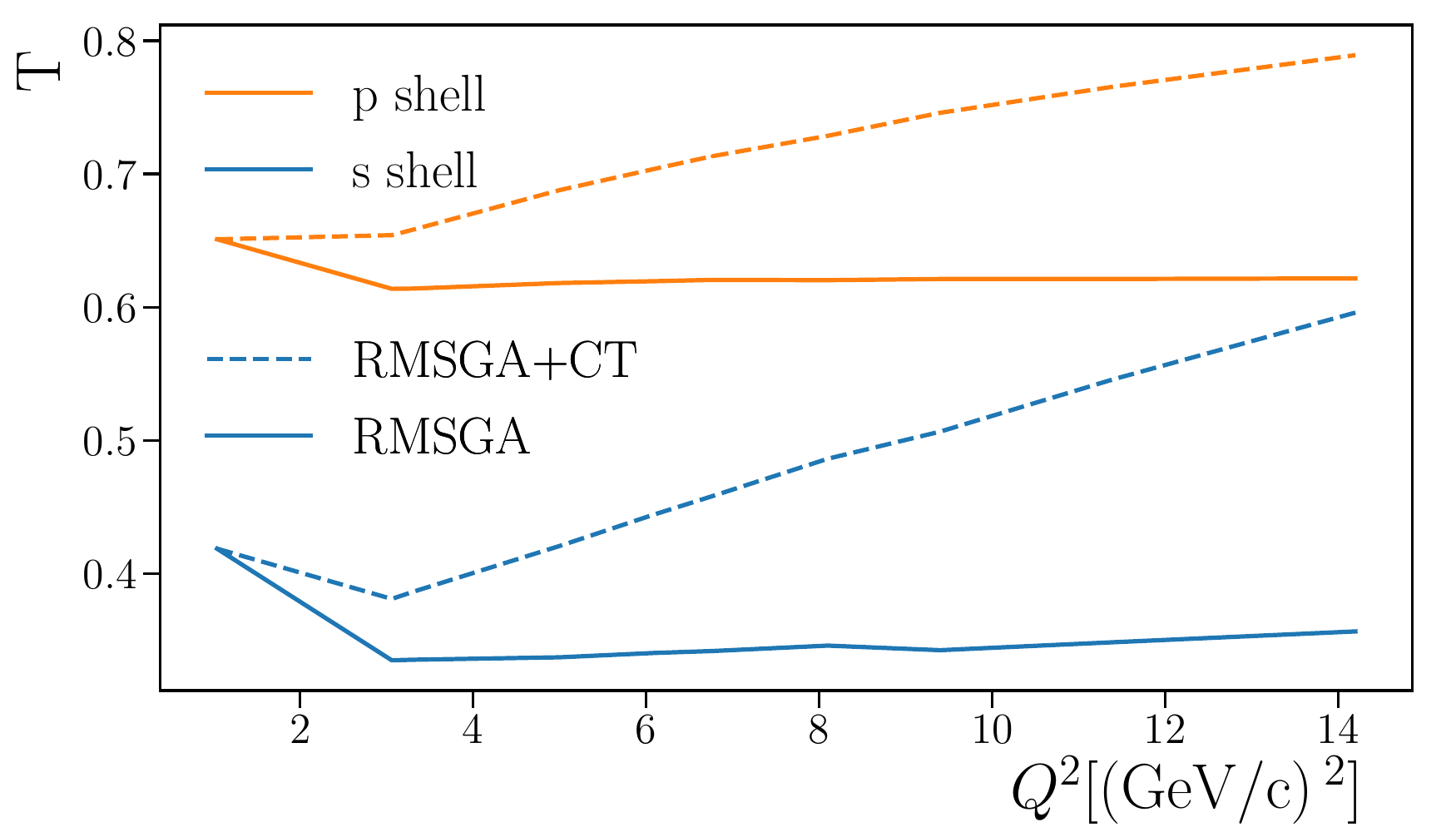}
\caption{Nuclear $(e,e'p)$ transparencies for the $s$-shell (blue) and $p$-shell (orange)  protons in $^{12}C$ versus $Q^2$ in quasielastic 
kinematics.  
Dashed curves include the CT effects. \label{fig:carbon_shells}}
\end{figure}

 Figure~\ref{fig:carbon_data} compares the 
RMSGA calculations for the $^{12}\text{C}(e,e'p)$ reaction with the data from several measurements, 
including the recent Hall C ones~\cite{HallC:2020ijh}.  This is essentially an update of 
 Figure~1 
 of 
Ref.~\cite{Cosyn:2013qe}, 
including the calculations for the new JLab12 data, which were included in the experimental publication~\cite{HallC:2020ijh}.  As reported in 
Ref.~\cite{HallC:2020ijh}, the data up to the highest measured $Q^2$ are in agreement with traditional Glauber calculations, including the RMSGA.  
As a further test of FSI models, more differential measurements could confront theory calculations in a more limited region of phase space.  For 
example, one could compare transparencies for proton knockout from the dense interior regions from the nucleus with those from peripheral regions.  
Thanks to the high-resolution spectrometers,
used in the Hall C measurement, the contributions from the $^{12}$C $s$-shell and $p$-shell proton 
knockout can be separated and these data are being prepared for publication.  
 In Figure~\ref{fig:carbon_shells}, new RMSGA predictions for the 
shell separated transparencies are shown.
 The calculations
 cover the $Q^2$ range of the world $A(e,e'p)$ transparency data, including calculations for the measured 
kinematics of Refs.~\cite{Makins:1994mm,Garrow:2001di,HallC:2020ijh}.  Both shells show a similar $Q^2$-dependence as the total transparency.  This 
is not surprising as this dependence of $T$ is largely governed by the momentum dependence of $\sigma_{NN}$, the total nucleon-nucleon cross section.  
Naturally the $s$-shell, having a smaller rms radius than the $p$-shell exhibits a smaller transparency.  The $p$-shell RMSGA transparencies 
excluding CT are about 75 \% larger than the $s$-shell ones and constant over the covered $Q^2$ range.  Given that the proton transparency aggregated 
over the different shells in $^{12}\text{C}$ of Figure~\ref{fig:carbon_data} did not exhibit any CT enhancement, 
 one anticipates
a similar observation 
for the separated shells.  For the sake of completeness we include the RMSGA transparencies with CT effects from the quantum diffusion model for 
proton knockout from the separated shells.  One observes that the $s$-shell transparency receives a larger enhancement from CT effects, again 
reflecting the fact that (on average) the FSI get reduced over a longer path length than for a $p$-shell proton.

To shed further light on the CT situation for the proton, measurements at higher momentum transfers at hadron beam facilities can provide additional 
data.  Additionally proton knockout can be measured in different kinematics; 
 see, e.g.,~\cite{Huber:2022wns}.  
 As a further  
theoretical study one can explore the parameter space of the quantum diffusion model in great detail and identify those parameter values that are 
compatible with the $^{12}\text{C}(e,e'p)$ measurements of Figure~\ref{fig:carbon_data}, but still lead to an observable change in the expected FSI 
in quasielastic deuteron breakup.  There, the kinematics can be tuned so that FSI are maximal, hence a CT signal would result in a bigger change in 
the transparency compared to the regular one.  Experimental measurements of deuteron breakup in these kinematics could then make clear if the current 
$(e,e'p)$ at high $Q^2$ would already be compatible with the formation of a point-like configuration during the knockout process.

 \vspace{2pt}

\textbf{Note added in proof:}  The Hall C data on the shell separated transparencies appeared in Ref.~\cite{Bhetuwal:2022wbe} after completion of this work. Note that the transparencies contained in~\cite{Bhetuwal:2022wbe} apply the overall correlation factor $c_A = 1.11$ to correct for nuclear short-range correlations.
The measured $p$-shell transparencies are in line with our RMSGA predictions, while the measured $s$-shell ones are about 25\% larger. This discrepancy
could be attributed to the fact that the measured $s$-shell
transparencies include data over a broad missing-energy range (from 20 to 80 MeV) whereby contributions from  reaction channels beyond $1s_{1/2}$ knockout cannot be excluded.  In addition, the results of Fig.~\ref{fig:carbon_shells} are obtained for a fixed missing energy corresponding with the energy-centroid of the single-particle strength.  A more detailed comparison between the data and the model for $s$-shell knockout requires a model for the wide energy distribution of this shell.
 \vspace{6pt}

\authorcontributions{Both authors contributed equally to all aspects.}

\funding{The work of W.C. is partially supported by the National Science Foundation under Award No. 2111442.} 

\dataavailability{Calculation results and computer codes are available upon request.}


\conflictsofinterest{The authors declare no conflict of interest.} 

\begin{adjustwidth}{-\extralength}{0cm}

\reftitle{References}

\end{adjustwidth}
\end{document}